\documentclass{webofc}
\usepackage[varg]{txfonts}   
\begin{document}
\title{On the nature of the $N^*$ and $\Delta$ resonances via coupled-channel dynamics}

\author{\firstname{Yu-Fei} \lastname{Wang}\inst{1}\fnsep\thanks{\email{yuf.wang@fz-juelich.de}}}

\institute{Institute for Advanced Simulation and J\"ulich Center for Hadron Physics, Forschungszentrum J\"ulich, 
52425 J\"ulich, Germany}

\abstract{This talk focuses on a recent work aiming at determining the composition of certain $N^*$ and $\Delta$ resonances, i.e. whether they are compact states formed directly by quarks and gluons, or composite generated from the meson-baryon interaction. The information of the resonance poles is provided by a comprehensive coupled-channel approach, the J\"{u}lich-Bonn model. Thirteen states that are significant in this approach are studied. Two criteria for each state are adopted in this paper, the comparison thereof roughly indicates the model uncertainties. It is found that the conclusions for eight resonances are relatively certain: $N(1535) \frac{1}{2}^-$, $N(1440) \frac{1}{2}^+$, $N(1710) \frac{1}{2}^+$, and $N(1520) \frac{3}{2}^-$ tend
to be composite; whereas $N(1650) \frac{1}{2}^-$, $N(1900) \frac{3}{2}^+$, $N(1680) \frac{5}{2}^+$, and
$\Delta(1600) \frac{3}{2}^+$ tend to be compact. }
\maketitle
The study of the inner structures of hadrons is always an important topic. Due to the non-pertubative nature of Quantum Chromodynamics (QCD), it is usually very difficult to figure out whether a hadron is formed directly by quark and gluons (elementary picture), or by residual interactions among hadrons (composite picture). The pioneer work traces back to the 1960ties, when Steven Weinberg related the structure of deuteron to the two-nucleon scattering length $a$ and effective range $r$~\cite{Weinberg1965}: 
\begin{equation}\label{Weinbergct}
    a=-\frac{2(1-Z)}{2-Z}R+\mathcal{O}(L)\ ,\quad
    r=-\frac{Z}{1-Z}R+\mathcal{O}(L)\ ;
\end{equation}
where $Z\in [0,1]$ is the probability to find a bare state in the deuteron (``elementariness''), $R\simeq 4.3$ fm is the binding radius, and $L\simeq m_\pi^{-1}$ is the interaction range. The experimental results for $a$ and $r$ support a rather small value of $Z$. Weinberg's criterion is model independent and compatible with the modern understanding of the effective field theories~\cite{Oller,Bira}. However, it can only be applied to $S$-wave stable bound states near the threshold. In recent decades more and more hadronic resonances are found, which are unstable and under complicated coupled-channel dynamics. Some of those states are believed to be hadronic molecules~\cite{FKGuo}. The generalizations of Weinberg's criterion are thus demanded. 

There are mainly three ways to establish the new criterion. First is the pole-counting rule~\cite{Morgan} stemming from the fact that typical quantum mechanical potentials can only generate one near-threshold pole in $S$-wave. Applications of this method can be found in Refs.~\cite{OZhang,LYDai,CMeng,QRGong,QFCao1,QFCao2,HChen}. Second is the spectral density function~\cite{SDF}: for resonances, the elementariness $Z$ in Eq.~\eqref{Weinbergct} disperses into a finite probability distribution $w(z)$ of the physical energy $z$. Further applications can be found in Refs.~\cite{QRGong,HChen,Kalashnikova1,Kalashnikova2,Baru2010,CHanhart1,CHanhart2}. The third method introduces the Gamow states~\cite{Gamow} to describe the resonances. The elementariness for a Gamow state is complex, but its magnitude can be interpreted by means of mathematical transformations~\cite{Guo} or naive measures~\cite{Oset,Sekihara1,Sekihara2}. Notice that model-independent analyses are always very difficult, especially for higher partial waves or broad states. Nevertheless, it is indicative to carry out a study with comprehensive models that are reliably constrained by large data sets. 

The J\"{u}lich-Bonn model~\cite{schuetz1994,schuetz1995,schuetz1998,Krehl2000,Gasparyan2003,Doering2009,Doering2011,Roenchen2013,Wang2022,Roenchen2014,Roenchen2015,Roenchen2018,Roenchen2022,Mai2021,Mai2022,Mai2023,Shen2018,ZLWang2022} is a very good candidate for the analyses of elementariness. It is a dynamical coupled-channel model for partial-wave analyses and resonance extraction in the energy region from $\pi N$ threshold to $2.3$ GeV, with the parameters determined by the fit to a worldwide collection of data. The amplitudes are calculated from Lippmann-Schwinger like equations so unitarity and analyticity~\cite{Doering2009} are respected. This talk is based on Ref.~\cite{mo}. We analyse thirteen significant $N^*$ and $\Delta$ resonances from the ``J{\"u}Bo2022'' solution~\cite{Roenchen2022}. Most of the poles are not in $S$-wave, so the pole-counting rule cannot be applied. Fortunately, the spectral density functions and the Gamow states can be directly extracted. 

To proceed, a solvable toy-model helps clarify the criteria. Consider the two-body problem with reduced mass $\mu$ and momentum $\mathbf{k}$. The orthogonal basis of the Hilbert space consists of one isolated bare state $|\psi_0\rangle$ and continuous states $|\psi(\mathbf{k})\rangle$. They are the eigenstates of the free Hamiltonian (with $E_0$ the bare energy): 
$\hat{H}_0=E_0|\psi_0\rangle\langle\psi_0|
+\int\frac{d^3 \mathbf{k}}{(2\pi)^3} \frac{k^2}{2\mu}|\psi(\mathbf{k})\rangle\langle\psi(\mathbf{k})|$. 
In addition, the interaction Hamiltonian only contains the $S$-wave coupling between the bare state and continuous states: 
$\hat{H}_I=\int \frac{d^3 \mathbf{k}}{(2\pi)^3}\ gF(k,\Lambda)
|\psi(\mathbf{k})\rangle\langle \psi_0 |+h.c.$, where $F(k,\Lambda)=\frac{\Lambda^2}{k^2+\Lambda^2}$ is the regulator with $\Lambda$ the cut-off. Any physical state with eigenenergy $\mathcal{E}$ is: 
\begin{equation}\label{Heigen}
    (\hat{H}_0+\hat{H}_I)|\Phi(\mathcal{E})\rangle=\mathcal{E}|\Phi(\mathcal{E})\rangle\ ,\quad
    |\Phi(\mathcal{E})\rangle=c_0(\mathcal{E})|\psi_0\rangle
    +\int\frac{d^3 \mathbf{k}}{(2\pi)^3}\chi(\mathbf{k},\mathcal{E})|\psi(\mathbf{k})\rangle\ .
\end{equation}
On the one hand, one can easily solve the bound state with $\mathcal{E}=-B<0$, by introducing the self-energy $\Sigma$: 
$\Sigma(\mathcal{E})\equiv
\int\frac{d^3\mathbf{k}}{(2\pi)^3}\frac{g^2F^2(k,\Lambda)}{\mathcal{E}-k^2/(2\mu)}$, 
$-B-E_0-\Sigma(-B)=0$ .
On the other hand, the ``elementariness'' is defined as the probability of finding the bare state in the bound state, and obtained by the normalization condition of bound states, 
$Z\equiv |\langle\psi_0|\Phi(-B)\rangle|^2=\frac{1}{1-\Sigma'(-B)}$. 
When the eigenenergy $\mathcal{E}>0$, the scattering states are solved by Lippmann-Schwinger equation. Performing the effective range expansion, one reproduces Weinberg's criterion Eq.~\eqref{Weinbergct}. 

To generalize the toy-model, one may introduce a lower channel $|\phi(\mathbf{k})\rangle$, with a very small coupling to the bare state. Then the bound state pole gains a small decay width $\Gamma_R$, with the resonance energy $E_R\simeq -B$. The projection of the scattering state to the bare state is the {\it spectral density function}: $w(\mathcal{E})\equiv |\langle\psi_0|\Phi(\mathcal{E})\rangle|^2$. When $\mathcal{E}\simeq E_R$, 
$w(\mathcal{E}\simeq E_R)=\frac{Z}{\pi}\frac{\Gamma_R/2}{(\mathcal{E}-E_R)^2+(\Gamma_R/2)^2}$. 
Therefore the spectral density function carries the information of the elementariness $Z$; when $\Gamma_R\to 0$, it reproduces the result for a bound state $w(\mathcal{E})=Z\delta(\mathcal{E}-E_R)$. If the state is too broad, the spectral density function loses this physical interpretation. All in all, one estimates the elementariness by collecting the spectral density function near the resonance energy: 
\begin{equation}\label{SDFZ}
    Z \simeq \frac{\int_{E_R-\Delta E}^{E_R+\Delta E}w(\mathcal{E})d\mathcal{E}}
	{\int_{E_R-\Delta E}^{E_R+\Delta E}BW(\mathcal{E})d\mathcal{E}}\ , \quad
	BW(\mathcal{E})\equiv \frac{1}{\pi}\frac{\Gamma_R/2}{(\mathcal{E}-E_R)^2+(\Gamma_R/2)^2}\ .
\end{equation}
In this work we choose $\Delta E=\Gamma_R$. At last, note that the spectral density function is also the K{\"a}ll{\'e}n-Lehmann spectral function, as the imaginary part of the propagator
\begin{equation}\label{ImD}
    D(\mathcal{E})\equiv \frac{1}{\mathcal{E}-E_0-\Sigma(\mathcal{E}+i0^+)}\ ,\quad 
    w(\mathcal{E})=-\frac{1}{\pi}\text{Im}D(\mathcal{E})\ .
\end{equation}

The discussions above do not introduce unphysical states. However, it is also feasible to define the resonance as a Gamow state. We reach the second (II) Riemann sheet of $D(\mathcal{E})$ by deforming the integral contour inside $\Sigma$, and then the resonance $\mathcal{E}_R$ is the pole of $D^{\text{II}}$. The deformation of the contour can also be applied to the definition of the state Eq.~\eqref{Heigen}, yielding the Gamow states $|\Phi(\mathcal{E}_R))$. Such states are the complex eigenstates of the Hamiltonian and cannot be normalized. An alternative normalization condition is defined by its conjugate pole at $\mathcal{E}_R^*$: $(\Phi(\mathcal{E}_R^*)|\Phi(\mathcal{E}_R))=1$, which leads to the elementariness of Gamow state: $Z_R=(1-\Sigma^{\text{II}\prime})^{-1}$. This is not a probability but a complex quantity. Moreover, as suggested in Ref.~\cite{Sekihara2}, one can also calculate the elementariness of the Gamow states via the off-shell residues $r(k)$ (``$C$'' labels the deformed contour; $X$ is called ``compositeness''): 
\begin{equation}\label{Xresoff}
    T\sim \frac{r^2(k)}{\mathcal{E}-\mathcal{E}_R}\ ,\quad 
    X=1-Z=\int_C\frac{d^3 \mathbf{k}}{(2\pi)^3}\frac{r^2(k)}{[\mathcal{E}_R-k^2/(2\mu)]^2}\ . 
\end{equation}

The J{\"u}lich-Bonn model differs to the toy-model in three aspects: i) relativistic kinematics; ii) coupled-channels and more bare states; iii) interaction potentials among the continuous states. For details of this model, see e.g. the supplemental material of Ref.~\cite{Wang2022}. The spectral density functions are still given by Eq.~\eqref{ImD}. For the $i$th bare state, one has a partial spectral density function $w_i(z)$ (with $z$ the centre-of-mass energy). The partial elementariness $Z_i$ can be calculated via Eq.~\eqref{SDFZ}, and then the ``total elementariness'' is estimated as $Z=1-\prod_i(1-Z_i)$. 

To estimate the uncertainties, we employ two other results as supplements to the spectral density functions directly from the model. First, we locally construct another spectral density function only from the residues and pole positions: 
\begin{equation}\label{SDFcst}
	w^{\text{lc}}(z)=-\frac{1}{\pi}\text{Im}\left[z-M_0-\sum_\kappa g^2_\kappa L_\kappa(z)\right]^{-1}\ ;
\end{equation}
where $\kappa$ is the channel index, $L$ is the loop function, and $M_0$, $g$ are parameters that reproduce the pole positions and residues in each channel. For some states this construction fails (no positive $g^2$ can be found), which indicates involved dynamics and larger model uncertainties. When this happens we just substitute the absolute values of normalized residues for the $g$'s with an extra constant width in $M_0$. Second, starting from the Gamow states, we calculate the ``partial compositeness'' $X_\kappa$ with the complex elementariness $Z=1-\sum_\kappa X_\kappa$, and then take the following naive measure~\cite{Sekihara2}: 
$\tilde{X}_\kappa\equiv\frac{|X_\kappa|}{\sum_\alpha|X_\alpha|+|Z|}$, 
$\tilde{Z}\equiv\frac{|Z|}{\sum_\alpha|X_\alpha|+|Z|}$. 

\begin{figure}[t]
\centering
\includegraphics[width=10cm,clip]{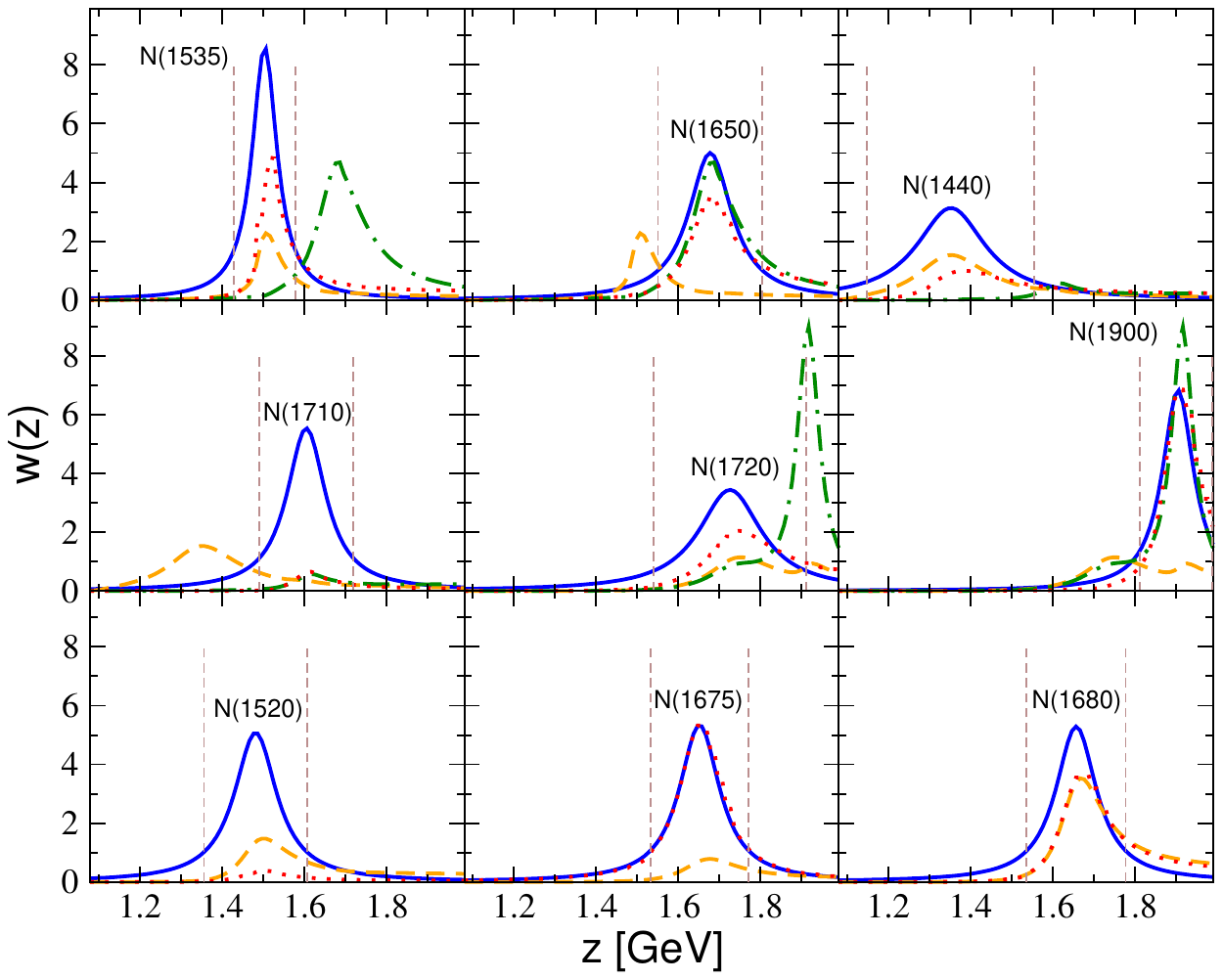}
\caption{The spectral density functions for the selected $N^*$ states. Blue solid line:
the Breit-Wigner denominator in Eq.~\eqref{SDFZ}. Orange dashed (green dash-dotted) line: the
1st (2nd) spectral density function from the model. Red dotted line: the locally constructed
function in Eq.~\eqref{SDFcst}. }
\label{fig:Nst}       
\end{figure}

\begin{figure}[t]
\centering
\includegraphics[width=8cm,clip]{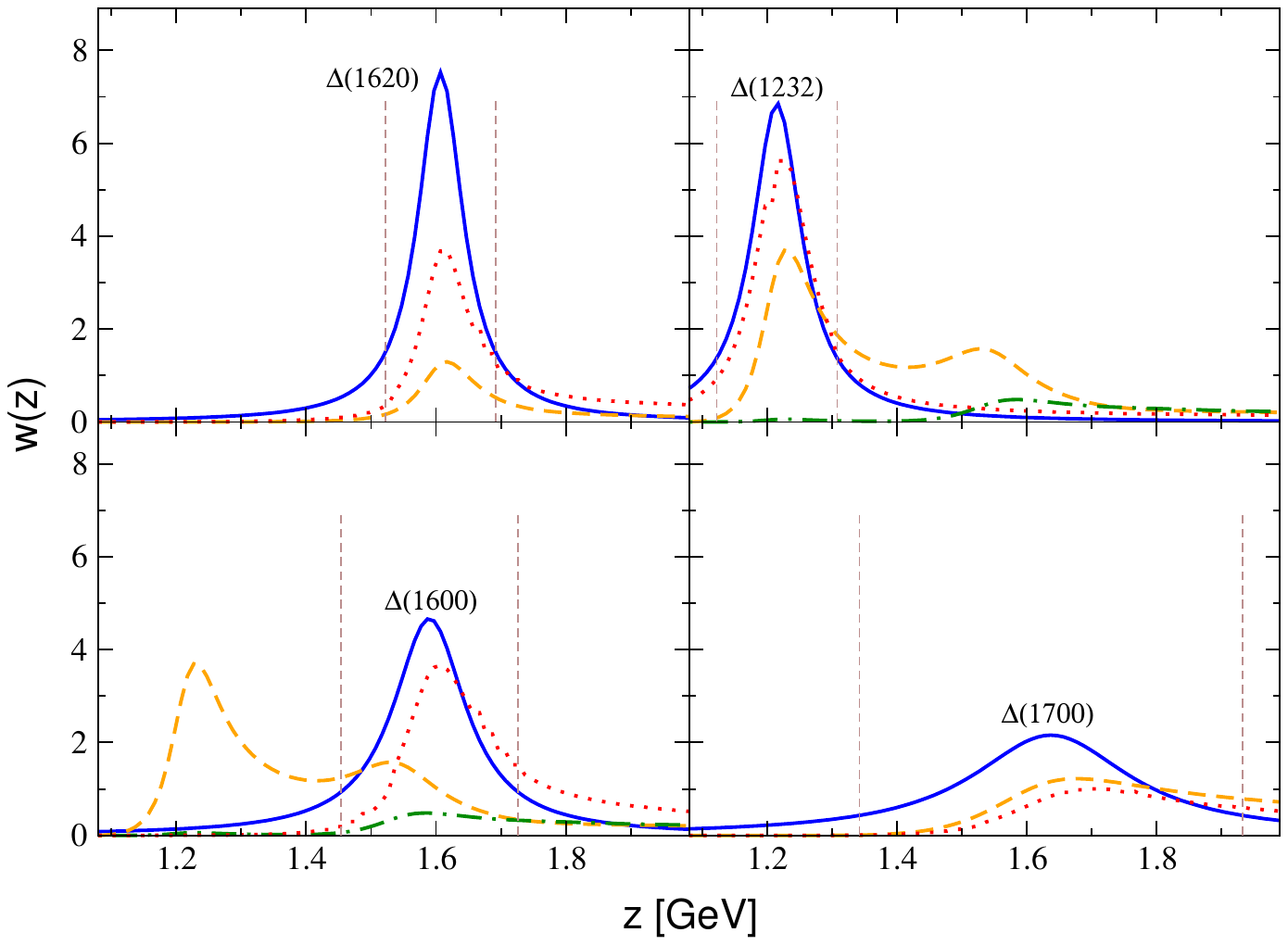}
\caption{The spectral density functions for the selected $\Delta$ states. Blue solid line:
the Breit-Wigner denominator in Eq.~\eqref{SDFZ}. Orange dashed (green dash-dotted) line: the
1st (2nd) spectral density function from the model. Red dotted line: the locally constructed
function in Eq.~\eqref{SDFcst}. }
\label{fig:D}       
\end{figure}

The analyses are based on Ref.~\cite{Roenchen2022}, the channel space of which is $\pi N$, $\pi\pi N$ (simulated by $\rho N$, $\sigma N$ and $\pi\Delta$), $\eta N$, $K\Lambda$, and $K\Sigma$. We choose the $N^*$'s with spin $J\leq 5/2$, and the $\Delta$'s with $J\leq 3/2$. Moreover, we exclude states with $\Gamma_R>300$ MeV. The results for thirteen states are listed in Tab.~\ref{tab}. The spectral density functions are in Figs.~\ref{fig:Nst} and \ref{fig:D}. 
\begin{table}
\centering
\footnotesize
\caption{The elementarinesses of the selected states. The label ``(F)'' means that the
local construction of Eq.~\eqref{SDFcst} has failed. }
\label{tab}
\begin{tabular}{lllll}
\hline
State & Pole position (MeV) & $\mathcal{Z}_{\rm tot}$ & $\mathcal{Z}^{\text{lc}}$ & $\tilde{Z}$\\
\hline
$N(1535)\,\frac{1}{2}^-$ & $1504-37i$ & $29.0\%$ & $50.8\%$ & $39.4\%$\\
$N(1650)\,\frac{1}{2}^-$ & $1678-64i$ & $92.8\%$ & $70.5\%$ & $8.5\%$\\
$N(1440)\,\frac{1}{2}^+$ & $1353-102i$ & $49.5\%$ & $31.5\%$ & $36.9\%$\\
$N(1710)\,\frac{1}{2}^+$ & $1605-58i$ & $20.6\%$ & $10.2\%$ & $40.3\%$\\
$N(1720)\,\frac{3}{2}^+$ & $1726-93i$ & $79.3\%$ & $62.5\%$ & $41.4\%$\\
$N(1900)\,\frac{3}{2}^+$ & $1905-47i$ & $100\%$ & $99.9\%$ & $38.5\%$\\
$N(1520)\,\frac{3}{2}^-$ & $1482-63i$ & $29.4\%$ & $7.2\%$ & $40.4\%$\\
$N(1675)\,\frac{5}{2}^-$ & $1652-60i$ & $16.6\%$ & (F) & $61.8\%$\\
$N(1680)\,\frac{5}{2}^+$ & $1657-60i$ & $67.9\%$ & $69.9\%$ & $55.0\%$\\
$\Delta(1620)\,\frac{1}{2}^-$ & $1607-42i$ & $18.9\%$ & $50.0\%$ & $69.4\%$\\
$\Delta(1232)\,\frac{3}{2}^+$ & $1215-46i$ & $53.8\%$ & (F) & $30.5\%$\\
$\Delta(1600)\,\frac{3}{2}^+$ & $1590-68i$ & $47.8\%$ & $77.5\%$ & $69.7\%$\\
$\Delta(1700)\,\frac{3}{2}^-$ & $1637-148i$ & $59.7\%$ & $44.9\%$ & $47.8\%$\\
\hline
\end{tabular}
\end{table}

It is found that at least two of the results suggest the composite picture for these four states: $N(1535) \frac{1}{2}^-$, $N(1440) \frac{1}{2}^+$, $N(1710) \frac{1}{2}^+$, $N(1520) \frac{3}{2}^-$. The complex compositeness of the Gamow state also gives indications of the dominant compositions: for $N^*(1535)$ $X_{\eta N}=35.8\%$, for $N^*(1440)$ $X_{\pi N}=59.0\%$, for $N^*(1710)$ $X_{\eta N}=44.9\%$, and for $N^*(1520)$ $X_{\pi\pi N}=43.7\%$. Nevertheless the compositions may be model dependent -- in this model the couplings of the $\sigma N$ channel to the bare states are all switched off; the $\omega N$ channel absent here can be important for $N^*(1535)$ as discussed in Ref.~\cite{Wang2022}. On the other hand, four states tend to be elementary: $N(1650) \frac{1}{2}^-$, $N(1900) \frac{3}{2}^+$, $N(1680) \frac{5}{2}^+$, and $\Delta(1600) \frac{3}{2}^+$. Note that in the literature the interpretations of some certain states are still open questions. For example it is  believed in some quark models that the $N^*(1440)$ is the radical excitation of the nucleon, but this picture is not supported in Ref.~\cite{Ulf1984}. Furthermore, this study has unfortunately not reached a conclusion for the $\Delta(1232)$ resonance, since there are some technical difficulties to handle the $\pi\Delta$ channel in the model. More discussions on this state can be found in Ref.~\cite{mo}. 

To summarize, this talk is on a recent study of the structures of the $N^*$ and $\Delta$ states based on the J{\"u}lich-Bonn model. It is suggested in this study that $N^*(1535)$, $N^*(1440)$, $N^*(1710)$, and $N^*(1520)$ are composite, while $N^*(1650)$, $N^*(1900)$, $N^*(1680)$, and $\Delta(1600)$ are elementary. In the future, this study can be extended when the $\omega N$ photoproduction reaction is analysed. We can also apply the method in this study to the hidden charm sector to figure out the structures of the $P_c$ states. 

I would like to thank the collaborators: Ulf-G.~Mei{\ss}ner, Deborah R{\"o}chen, and Chao-Wei Shen. The computing time granted by the JARA Vergabegremium and provided on the JARA Partition part of the supercomputer JURECA~\cite{JU} at Forschungszentrum J{\"u}lich is acknowledged. This work is supported by the NSFC and the Deutsche Forschungsgemeinschaft (DFG, German Research Foundation) through the funds provided to the Sino-German Collaborative Research Center TRR110 “Symmetries and the Emergence of Structure in QCD” (NSFC Grant No. 12070131001, DFG Project-ID 196253076-TRR 110). Further support by the CAS through a President’s International Fellowship Initiative (PIFI) (Grant No. 2018DM0034) and by the VolkswagenStiftung (Grant No. 93562) is acknowledged. This work is also supported by the MKW NRW under the funding code NW21-024-A. 

\end{document}